\documentclass[iop,revtex4]{emulateapj}
\usepackage{graphicx}
\usepackage[colorlinks,urlcolor=red,citecolor=blue,linkcolor=cyan]{hyperref} 
\newcommand{\icarus}{Icarus}
\def\shk{$S_{HK}$~}
\slugcomment{Accepted for publication in ApJ}
\begin{document}

\shorttitle{A 12-Year Activity Cycle for HD 219134}
\shortauthors{Johnson et al.}

\title{A 12-Year Activity Cycle for the Nearby Planet Host Star HD 219134}

\author{Marshall C. Johnson\altaffilmark{1}, Michael Endl\altaffilmark{1}, William D. Cochran\altaffilmark{1}, Stefano Meschiari\altaffilmark{1}, Paul Robertson\altaffilmark{2,3,4}, \\ Phillip J. MacQueen\altaffilmark{1}, Erik J. Brugamyer\altaffilmark{1}, Caroline Caldwell\altaffilmark{5}, Artie P. Hatzes\altaffilmark{6}, Ivan Ram\'irez\altaffilmark{1}, \\ and Robert A. Wittenmyer\altaffilmark{7,8,9}}

\altaffiltext{1}{Department of Astronomy and McDonald Observatory, University of Texas at Austin, 2515 Speedway, Stop C1400, Austin, TX 78712, USA; mjohnson@astro.as.utexas.edu}
\altaffiltext{2}{NASA Sagan Fellow}
\altaffiltext{3}{Department of Astronomy and Astrophysics, The Pennsylvania State University, USA}
\altaffiltext{4}{Center for Exoplanets \& Habitable Worlds, The Pennsylvania State University, USA}
\altaffiltext{5}{Astrophysics Research Institute, Liverpool John Moores University, 146 Brownlow Hill, Liverpool, L3 5RF, UK}
\altaffiltext{6}{Th\"uringer Landessternwarte Tautenburg, Sternwarte 5, D-07778 Tautenberg, Germany}
\altaffiltext{7}{School of Physics, University of New South Wales, Sydney 2052, Australia}
\altaffiltext{8}{Australian Centre for Astrobiology, University of New South Wales, Sydney 2052, Australia}
\altaffiltext{9}{Computational Engineering and Science Research Centre, University of Southern Queensland, Toowoomba, Queensland 4350, Australia}

\begin{abstract}

The nearby (6.5 pc) star HD 219134 was recently shown by \cite{Motalebi15} and \cite{Vogt15} to host several planets, the innermost of which is transiting. We present twenty-seven years of radial velocity observations of this star from the McDonald Observatory Planet Search program, and nineteen years of stellar activity data. 
We detect a long-period activity cycle measured in the Ca \textsc{ii} $S_{HK}$ index, with a period of $4230 \pm 100$ days (11.7 years), very similar to the 11-year Solar activity cycle. Although the period of the Saturn-mass planet HD 219134 h is close to half that of the activity cycle, we argue that it is not an artifact due to stellar activity. We also find a significant periodicity in the $S_{HK}$ data due to stellar rotation with a period of 22.8 days. This is identical to the period of planet f identified by \cite{Vogt15}, suggesting that this radial velocity signal might be caused by rotational modulation of stellar activity rather than a planet. Analysis of our radial velocities allows us to detect the long-period planet HD 219134 h and the transiting super-Earth HD 219134 b. Finally, we use our long time baseline to constrain the presence of longer-period planets in the system, excluding to $1\sigma$ objects with $M\sin i>0.36 M_J$ at 12 years (corresponding to the orbital period of Jupiter) and $M\sin i>0.72 M_J$ at a period of 16.4 years (assuming a circular orbit for an outer companion). 

\end{abstract}


\section{Introduction}

HD 219134 (aka HR 8832, HIP 114622) is a nearby ($d=6.5$ pc) K3V star. As one of the nearest, brightest K dwarfs, it has long been a target of radial velocity (RV) planet surveys. Indeed, it was one of the targets of the first RV planet search, that of \cite{Walker95}, who began observing the star in 1980. It was also one of the original targets of our own McDonald Observatory Planet Search, with observations beginning in 1988 \citep{CochranHatzes93}.

Despite the long history of RV observations of HD 219134, it was not until the advent of modern high-precision, high-stability spectrographs (with long-term internal precision of $\sim1$ m s$^{-1}$) that planets have actually been detected around this star. \cite{Motalebi15} (hereafter M15) presented radial velocity observations of this star with the HARPS-N spectrograph. Using these data, they detected four planets, HD 219134 b, c, d, and e, with periods of 3.094, 6.765, 46.78 and 1842 days and $M\sin i$ values of 4.46, 2.67, 8.67 and 71 $M_{\oplus}$, respectively. The orbital period of their outer planet, HD 219134 e, is longer than the $\sim1100$ day span of their observations, and so its orbital period was not well constrained; formally, they found $P=1842_{-292}^{+4199}$ days. They also detected one transit of the innermost planet, HD 219134 b, using the Spitzer Space Telescope. 
This is the nearest transiting exoplanet discovered to date, and, with $V=5.57$, HD 219134 is the brightest star known to host a transiting exoplanet. 

Meanwhile, \cite{Vogt15} (hereafter V15) analyzed data from Keck/HIRES and the Automated Planet Finder (APF) on this system, and found six planets. They detected the three inner planets found by M15, plus two additional super-Earths (HD 219134 f and g) with periods of 22.8 and 94.2 days, respectively. They found a period of 2247 days for the outer Saturn-mass planet, significantly larger than that available at the time from M15; they therefore labeled this planet HD 219134 h. For the remainder of this paper we will refer to the outer planet as HD 219134 h rather than e (except when referring directly to M15), as the parameters we measured from our data more closely match the V15 values. Given the very large uncertainty on the period of the outer planet found by M15, however, the periods found by these two works are identical to within $1\sigma$.

V15 and M15 disagreed on the rotation period of HD 219134. V15 estimated a period of $\sim20$ days based upon the measured $v\sin i$ of the star \citep[1.8 km s$^{-1}$, from][]{ValentiFischer05}, while M15 found a period of 42.3 days from periodogram analysis of their stellar activity measurements ($\log R^\prime_{HK}$, derived from the Ca~\textsc{ii} H and K lines, and the cross correlation function bisector span and FWHM). M15 also found a smaller value of the $v\sin i$ of $0.4 \pm 0.5$ km s$^{-1}$ with the higher resolution of HARPS-N. The rotation period is important as stellar activity correlated with the rotation period--due to spots and active areas moving in and out of view--can create RV signals that masquerade as planets \citep[e.g.,][]{robertson14}. A more accurate measurement of the rotation period for HD 219134 could thus help to confirm the planetary status of, or refute as a false positive, the 22.8 (planet f) and 46.78 day (planet d) RV signals. 

HD 219134 is important as it is one of the nearest, brightest stars with a planetary system, as well as the nearest, brightest star to host a known transiting exoplanet. Future observations to further characterize this system will therefore be important. In particular, with multiple short-period super-Earths, this system appears to be a nearby analog to the many systems of closely-packed transiting super-Earths found by {\it Kepler} \citep[e.g.,][]{Rowe14}. The brightness of HD 219134 as compared to the {\it Kepler} sample offers an opportunity to pursue detailed characterization of the system that is difficult for most {\it Kepler} systems.

Recent theoretical work has suggested that the presence, or lack thereof, of long-period giant planets could affect the formation of such systems. \cite{BatyginLaughlin15} argued that the migration of Jupiter within our own solar system might have disrupted a massive primordial inner protoplanetary disk that could have formed multiple short-period super-Earths; they predicted that systems like the {\it Kepler} short-period multiple systems should typically lack long-period giant planets. A related question is, how common are planetary systems broadly similar in architecture to our solar system, with small close-in planets and more distant giant planets? We can begin to answer these questions in the near future through the combination of searches for short-period super-Earths and data from the long-term RV programs that have been monitoring many bright FGK stars for well over a decade. Super-Earths can be found with either high-precision RV observations or space-based transit searches. Such high-precision RV surveys include those being undertaken currently with HARPS \citep[e.g.,][]{diaz15}, HARPS-N (M15), APF \citep{Vogt14}, and CHIRON \citep{Tokovinin13}, and in the near future with MINERVA \citep{Swift15}, CARMENES \citep{Quirrenbach14}, ESPRESSO \citep{Megevand14}, and SPIRou \citep{Artigau14}. The major upcoming space-based transit survey is that of TESS \citep{Ricker15}. Long-term RV programs include the McDonald Observatory Planet Search \citep[e.g.,][]{Endl15}, the Anglo-Australian Planet Search \citep[e.g.,][]{Jones10}, the Lick-Carnegie Exoplanet Survey \citep[e.g.,][]{Rowan15}, the CORALIE planet search \citep{Marmier13}, and the planet search at ESO \citep[e.g.,][]{Zechmeister13}. Long-period giant planets will also be found by {\it Gaia}, which will produce a huge sample of astrometrically-detected planets \citep{Perryman14}. While most of the {\it Kepler} sample is too faint to have been observed previously by long-term RV surveys \citep[e.g.,][]{Coughlin15}, {\it Gaia} will be able to astrometrically detect long-period planets around many of these stars. Our own McDonald Observatory Planet Search program now has a baseline of 12-15 years for $\sim200$ FGKM stars, and a handful of stars also have lower-precision observations dating back more than 25 years. HD 219134 is one of these stars, and here we present an analysis of our radial velocity observations of this star, as well as our data on the stellar activity. 

\section{Observations}

\subsection{McDonald Observations}

We observed HD 219134 with the coud\'e spectrograph on the 2.7 m Harlan J.~Smith Telescope (HJST) at McDonald Observatory. 
We obtained 295 spectra of HD 219134 between 1988 July 26 UT and 2015 October 16 UT, using three different spectrograph formats. During Phase I of our program (1988 July 26 to 1995 July 20), we obtained 30 spectra using the telluric O$_2$ band at 6300 \AA~as a velocity reference. Phase II (1990 October 14 to 1997 November 16), during which we obtained 34 spectra, used a standard I$_2$ absorption cell as the velocity standard. We used spectrograph configurations with a resolving power of $R\sim210,000$ for Phase I and II; only a single spectral order was observed. Finally, Phase III (1998 July 16 to present, 231 spectra) continues to use the I$_2$ cell but uses the Robert G.~Tull Spectrograph's TS23 \citep{Tull95} configuration. This is a cross-dispersed \'echelle spectrograph with a spectral resolving power of $R=60,000$ and coverage from 3750 \AA~to 10200~\AA, including complete coverage blueward of 5691 \AA. Continual incremental improvements to the instrument and observing procedures over this time have increased the radial velocity precision. See \cite{CochranHatzes93} and \cite{HatzesCochran93} for more detail on Phases I and II, and \cite{Hatzes03} for Phase III.

In order to monitor the stellar activity, we measure the Ca S-index ($S_{HK}$), derived from the Ca~\textsc{ii} H and K lines \citep{Soderblom91,Baliunas95,Paulson02}. We could only perform this measurement for the Phase III data; the single-order format of the Phase I and II data does not include the H and K lines.
Our HJST RV and \shk measurements are listed in Table~\ref{HJSTdata}.

\subsection{Keck Observations}

We also observed HD 219134 with the HIRES spectrograph \citep{Vogt94} on the Keck I telescope; ironically, it served as a radial velocity standard for several of our programs due to its low RV variation. These programs included an RV planet search in the Hyades \citep[e.g.,][]{Cochran02} and the CoRoT NASA Key Science Project. For the latter project, we observed HD 219134 while the CoRoT field was not observable. We obtained 72 spectra between 1996 October 6 and 2000 December 3 UT, and 288 more between 2005 December 9 and 2012 January 11 UT, for a total of 360 spectra. The major difference between these datasets is that the earlier observations used an $2048\times2048$ Tektronics CCD, whereas the newer dataset used a $3\times1$ mosaic of $2048\times4096$ CCDs. The time sampling of the these datasets is rather uneven, with all of the 360 spectra being obtained during only fifteen observing runs, nine between 1996 and 2000 and six between 2005 and 2012. 

We also measured \shk from our Keck spectra. For the data obtained from 1996 to 2000 this required special care, as the HIRES format used during these years caused order overlap in the region of the Ca H and K lines for certain slit lengths, which was not always accounted for in the calibration sequence. This necessitated careful attention to the removal of scattered light during the data reduction process and likely results in systematics in the measurements. Our Keck RVs and \shk measurements are listed in Table \ref{Keckdata}.

\subsection{Data Reduction}

We reduced the data and extracted the spectra using a pipeline based on standard IRAF tasks. We measured the radial velocities of the spectra using the \textsc{Austral} I$_2$ cell reduction code \citep{Endl00}. See \cite{Endl15} for more detail on the reduction process.

\begin{deluxetable}{lcrccc}
\tabletypesize{\scriptsize}
\tablecolumns{6}
\tablewidth{0pt}
\tablecaption{HJST RV and S-index Data \label{HJSTdata}}
\tablehead{
\colhead{Row} & \colhead{BJD} & \colhead{dRV} & \colhead{$\sigma_{\mathrm{dRV}}$} & \colhead{$S_{HK}$} & \colhead{$\sigma_S$} \\
\colhead{ } & \colhead{ } & \colhead{(m s$^{-1}$)} & \colhead{(m s$^{-1}$)} & \colhead{ } & \colhead{ }
}

\startdata
   & Phase I       &    \ldots   &  \ldots    &     \ldots     &     \ldots   \\
1 & 2447368.95340 & -1.3 & \ldots & \ldots & \ldots \\  
2 & 2447369.91300 & 21.7 & \ldots & \ldots & \ldots \\  
3 & 2447429.84020 & -0.1 & \ldots & \ldots & \ldots \\  
4 & 2447459.73290 & -1.6 & \ldots & \ldots & \ldots \\  
5 & 2447460.74170 & -47.4 & \ldots & \ldots & \ldots \\  
6 & 2447495.71880 & 9.2 & \ldots & \ldots & \ldots \\  
7 & 2447496.69230 & 1.3 & \ldots & \ldots & \ldots \\  
8 & 2447516.65010 & 10.3 & \ldots & \ldots & \ldots \\  
9 & 2447517.63920 & 12.5 & \ldots & \ldots & \ldots \\  
10 & 2447551.55820 & -0.9 & \ldots & \ldots & \ldots \\  
\hline
   & Phase II      &    \ldots      &   \ldots   &    \ldots   &  \ldots      \\
31 & 2448178.72267 & 44.5 & \ldots & \ldots & \ldots \\  
32 & 2448223.56801 & -26.6 & \ldots & \ldots & \ldots \\  
33 & 2448259.63987 & 16.4 & \ldots & \ldots & \ldots \\  
34 & 2448485.92949 & 23.7 & \ldots & \ldots & \ldots \\  
35 & 2448524.77848 & 35.8 & \ldots & \ldots & \ldots \\  
36 & 2448555.74440 & -16.3 & \ldots & \ldots & \ldots \\  
37 & 2448825.88446 & 5.7 & \ldots & \ldots & \ldots \\  
38 & 2448853.88667 & 28.3 & \ldots & \ldots & \ldots \\  
39 & 2448882.79778 & -23.8 & \ldots & \ldots & \ldots \\  
40 & 2448902.74426 & -7.5 & \ldots & \ldots & \ldots \\
\hline
   & Phase III     &    \ldots  &  \ldots    &    \ldots      &    \ldots    \\
65 & 2451010.86514 & 1.7 & 3.7 & 0.249 & 0.020 \\  
66 & 2451065.83986 & -11.5 & 4.3 & 0.263 & 0.021 \\  
67 & 2451151.66780 & -0.9 & 4.5 & 0.261 & 0.019 \\  
68 & 2451211.58199 & 0.4 & 5.1 & 0.216 & 0.019 \\  
69 & 2451417.92148 & -2.0 & 4.1 & 0.231 & 0.021 \\  
70 & 2451451.83583 & 5.2 & 4.2 & 0.228 & 0.020 \\  
71 & 2451502.68841 & -1.0 & 4.3 & 0.244 & 0.022 \\  
72 & 2451505.60059 & 22.7 & 5.8 & 0.225 & 0.020 \\  
73 & 2451530.68963 & -10.6 & 5.0 & 0.201 & 0.019 \\  
74 & 2451557.61677 & -0.4 & 4.7 & 0.198 & 0.017

\enddata

\tablecomments{Table \ref{HJSTdata} is published in its entirety in the electronic edition of ApJ. A portion is shown here for guidance regarding its form and content. The velocities have been shifted such that the mean velocity of each dataset is 0. The internal uncertainties that we calculated for the Phase I and II data are not reliable and so are not quoted. The standard deviation of the measurements are 23~m~s$^{-1}$ and 21 m s$^{-1}$ for the Phase I and II data, respectively. Note that although both Phase I and II data were obtained between 1990 and 1995, we list these data in separate portions of the table rather than interspersing them.}

\end{deluxetable}

\begin{deluxetable}{lcrccc}
\tabletypesize{\scriptsize}
\tablecolumns{6}
\tablewidth{0pt}
\tablecaption{Keck RV and S-index Data \label{Keckdata}}
\tablehead{
\colhead{Row} & \colhead{BJD} & \colhead{dRV} & \colhead{$\sigma_{\mathrm{dRV}}$} & \colhead{$S_{HK}$} & \colhead{$\sigma_S$} \\
\colhead{ } & \colhead{ } & \colhead{(m s$^{-1}$)} & \colhead{(m s$^{-1}$)} & \colhead{ } & \colhead{ }
}

\startdata
   & Old CCD       &  \ldots     &   \ldots   &   \ldots       &   \ldots     \\
1 & 2450362.92346 & 2.7 & 4.4 & \ldots & \ldots \\  
2 & 2450362.92568 & -3.7 & 4.2 & 0.313 & 0.017 \\  
3 & 2450362.92750 & -5.4 & 4.2 & 0.318 & 0.018 \\  
4 & 2450362.92983 & -7.0 & 4.1 & 0.302 & 0.017 \\  
5 & 2450362.93161 & -5.2 & 4.3 & 0.318 & 0.017 \\  
6 & 2450362.93339 & -3.4 & 4.5 & 0.322 & 0.017 \\  
7 & 2450362.93517 & -3.9 & 4.2 & 0.320 & 0.017 \\  
8 & 2450362.93697 & -4.7 & 4.2 & 0.329 & 0.017 \\  
9 & 2450362.93865 & -5.5 & 4.4 & 0.299 & 0.016 \\  
10 & 2450362.94052 & -6.9 & 4.2 & 0.322 & 0.017 \\  
\hline
   & New CCD      &    \ldots    &   \ldots   &    \ldots      &    \ldots    \\
73 & 2453713.68535 & 10.5 & 4.2 & 0.244 & 0.006 \\  
74 & 2453713.68673 & 11.5 & 4.3 & 0.242 & 0.006 \\  
75 & 2453713.75681 & 2.3 & 4.1 & 0.237 & 0.006 \\  
76 & 2453713.81749 & 11.3 & 4.2 & 0.243 & 0.006 \\  
77 & 2453713.81847 & 11.0 & 4.3 & 0.242 & 0.006 \\  
78 & 2453714.71269 & 10.8 & 4.4 & 0.243 & 0.006 \\  
79 & 2453714.71367 & 16.7 & 4.6 & 0.243 & 0.005 \\  
80 & 2453714.77784 & 17.8 & 4.6 & 0.240 & 0.005 \\  
81 & 2453714.77896 & 15.9 & 4.6 & 0.232 & 0.006 \\  
82 & 2453714.86941 & 21.9 & 4.5 & 0.216 & 0.005
\enddata

\tablecomments{Table \ref{Keckdata} is published in its entirety in the electronic edition of ApJ. A portion is shown here for guidance regarding its form and content. The velocities have been shifted such that the mean velocity of each dataset is 0.}

\end{deluxetable}

\section{Stellar Activity and $S_{HK}$}
\subsection{Stellar Activity Cycle}

\begin{figure}
\epsscale{1.25}
\plotone{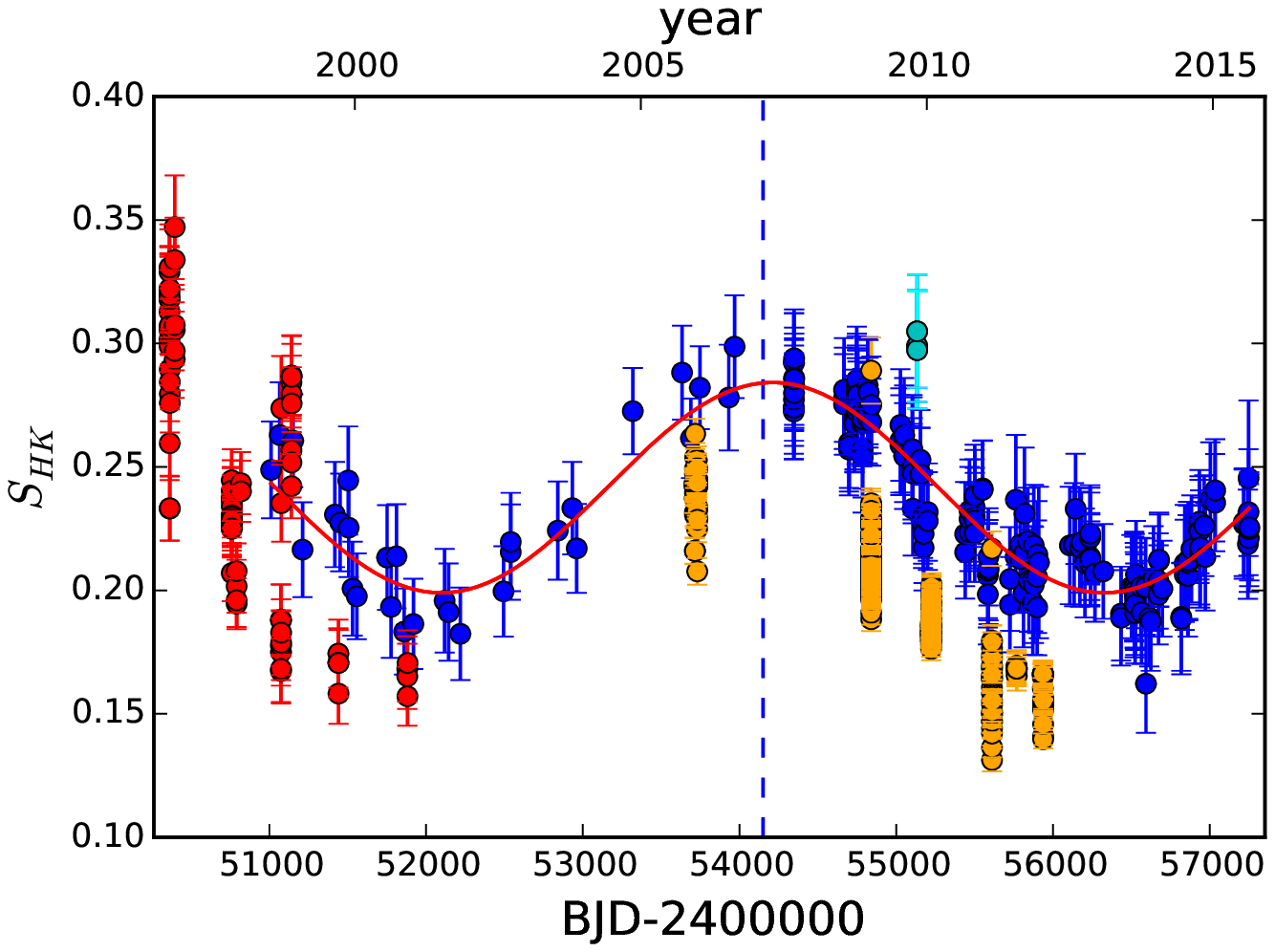}
\plotone{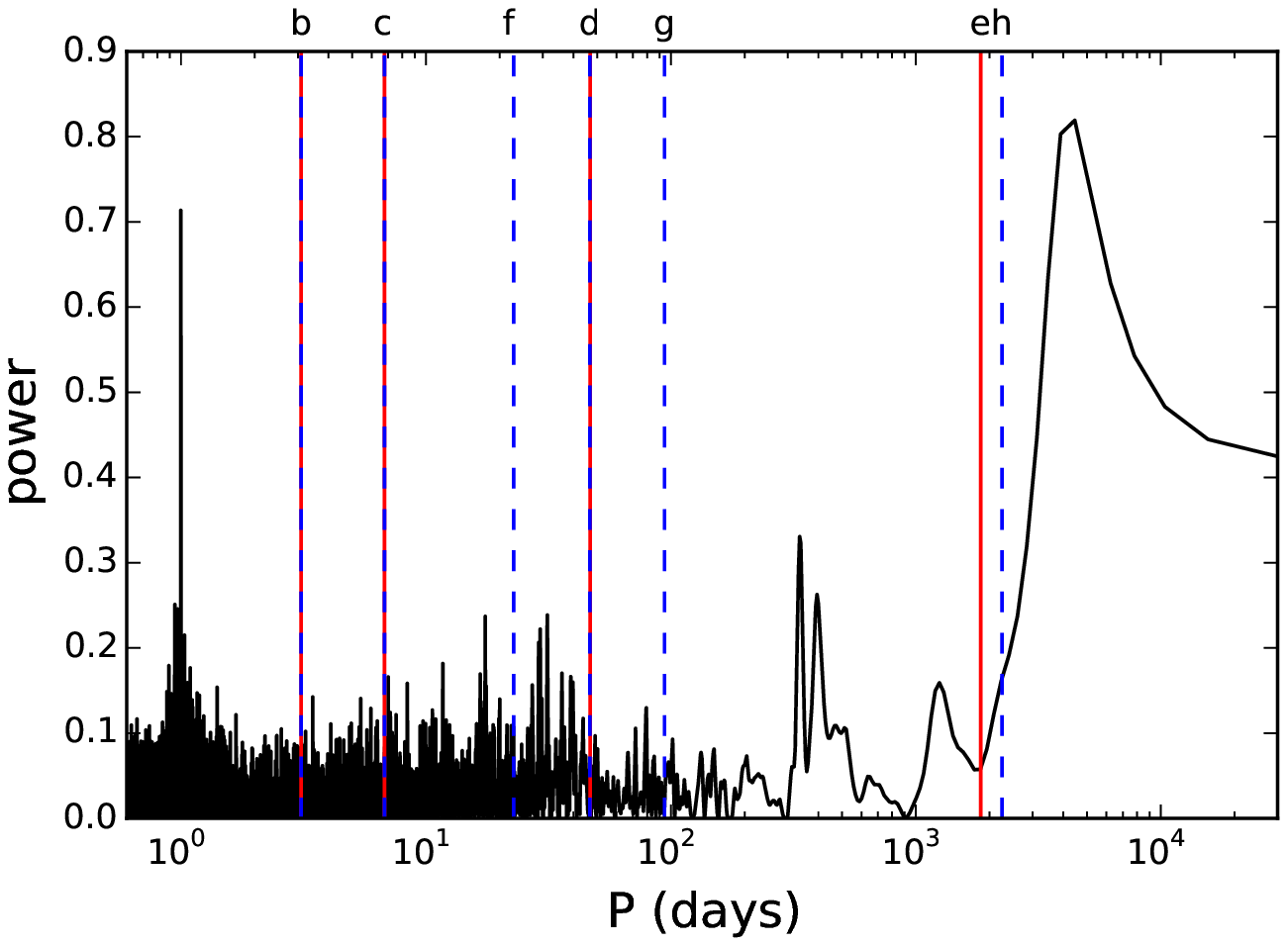}
\caption{Top panel: our \shk measurements, with TS23 data shown in blue, and Keck data from the old and new CCDs shown in red and orange, respectively. Note the systematic offset between the TS23 and Keck data, and the large scatter between adjacent observing runs in the Keck old CCD data. The best-fit sinusoidal activity cycle model (produced by fitting to the TS23 data only) is overplotted in red. A vertical blue dashed line marks the beginning of the 2007 observing season. Due to our higher post-2007 observing cadence, only data from after this point were used to generate the periodograms shown in the bottom two panels of Fig.~\ref{fig:rot}. Three TS23 points with abnormally high \shk values (possibly due to a flare) are shown in light blue; these were excluded from the bottom panel of Fig.~\ref{fig:rot}. Additionally, one extremely deviant TS23 datapoint has been excluded from this plot. Lower panel: Generalized Lomb-Scargle periodogram of the TS23 $S_{HK}$ measurements. Vertical red and dashed blue lines mark the periods of the planets found by M15 and V15, respectively, while the names of the planets are marked along the top of the panel. Note that the long-period Saturn-mass planet was denoted planet e by M15 and planet h by V15. \label{smw+model}}
\end{figure}

Our $S_{HK}$ data are shown in Fig.~\ref{smw+model}. Even by eye, it is apparent that there is an approximately sinusoidal long-term variation in $S_{HK}$. The generalized Lomb-Scargle (GLS) periodogram of these data shows a very strong peak at $P\sim4000$ days (lower panel of Fig.~\ref{smw+model}). There is a systematic offset between the Keck and the TS23 (Phase III) data (upper panel of Fig.~\ref{smw+model}). There also appear to be significant systematic offsets between different observing runs with the old Keck CCD, and moreover a large amount of scatter within these data.

We fit a simple sinusoidal model for the activity cycle to the data with a Markov chain Monte Carlo (MCMC) using the \texttt{emcee} package \citep{emcee}. We did not include the Keck old CCD data due to the apparent systematic offsets between successive observing runs.  
We attempted to include the newer Keck data with a multiplicative offset to bring it into agreement with the TS23 $S_{HK}$ measurements. A multiplicative offset is appropriate because the offsets between the datasets are due to differing spectrograph optical throughput and detector quantum efficiency. Including the Keck data, however, irrespective of what binning was used, always skewed the best-fit sinusoid away from that obvious from the TS23 data alone. Therefore, in order to measure the activity cycle parameters we used only the TS23 data. We fit for four parameters: the activity cycle period $P_{SHK}$, the amplitude $A_{SHK}$, the mean $S_{HK}$ level $<S_{HK}>$, and the epoch of minimum activity $t_{SHK,\mathrm{min}}$. 
We obtained a best-fit activity cycle period of $P_{SHK}=4230 \pm 100$ days. See Table~\ref{actpars} for the full list of best-fit parameters. The best-fit model, along with the data, is shown in the upper panel of Fig.~\ref{smw+model}.

\begin{deluxetable}{lcc}
\tabletypesize{\scriptsize}
\tablecolumns{3}
\tablewidth{0pt}
\tablecaption{Activity Fit Parameters \label{actpars}}
\tablehead{
\colhead{Parameter} & \colhead{Value} & \colhead{$1\sigma$ Uncertainty}
}

\startdata
Activity Cycle & & \\
$A_{SHK}$ & 0.0426 & 0.0024 \\
$<S_{HK}>$ & 0.2416 & 0.0018 \\
$P_{SHK}$ (days) & 4230 & 100 \\
$t_{SHK,\mathrm{min}}$ (BJD) & 2452096 & 81 \\
$<\log R^\prime_{HK}>$ & -4.89 & \ldots 
\enddata

\tablecomments{Best-fit parameters for a sinusoidal fit to the TS23 $S_{HK}$ measurements.
See the text for details.}

\end{deluxetable}

The best-fit activity cycle period of 11.6 years is very similar in period to the 11-year solar cycle. HD 219134 thus joins a growing number of stars with long-term activity cycles detected by RV planet search programs \citep[e.g.,][]{Lovis11,Robertson13Mact,Endl15}.

Both V15 and M15 also monitored the stellar activity using the Ca~\textsc{ii} H and K lines, the former measuring this as $S_{HK}$ and the latter as $\log R^\prime_{HK}$. Neither of these works, however, were able to detect the full activity cycle. The M15 dataset spans $\sim1100$ days near the most recent minimum of the activity cycle, and this time span was insufficient to measure the 4200-day period of the cycle. V15, on the other hand, presented nearly 4000 days of reasonably high cadence data, plus three datapoints $\sim3000$ days earlier, and yet did not detect the activity cycle (although their periodogram of these data shows a peak at 3290 days). Their lack of detection of the activity cycle appears to have been simply due to chance misfortune. Their data covered much of one cycle, but began only slightly before one activity maximum and ended slightly after the next activity minimum. This apparently conspired with possible systematic offsets in several sets of $S_{HK}$ measurements near the activity maximum, and a possible systematic offset between the HIRES and APF activity measurements, to give the appearance of a linear trend (see the top panel of Fig.~5 of V15). Furthermore, the three HIRES $S_{HK}$ measurements from 1996 were unluckily located near the previous activity maximum, further promoting the appearance of a linear trend.

\subsection{Stellar Rotation Period}
\label{Paulswork}

Based on analysis of the HARPS-N time series activity indicators, M15 concluded that HD 219134 has a rotation period of 42.3 days.  
This is inconsistent with the results of V15, who determined the rotation period must be closer to 20 days based on the measurement of $v \sin i=1.8$~km~s$^{-1}$ from \cite{ValentiFischer05}. On the other hand, the value of $v\sin i=0.4 \pm 0.5$~km~s$^{-1}$ found by M15 (using the very high resolution of HARPS-N) would predict a rotation period of $\sim98$ days, with a 1$\sigma$ lower limit of 44 days, assuming $\sin i\sim1$. The rotation period is of particular significance for HD 219134, as several of its recently-discovered planets have periods close to candidate rotation periods, introducing the possibility that one or more of the RV signals interpreted as exoplanets may instead be caused by stellar magnetic activity. M15 noted that the 46.8-day period of HD 219134 d is close to their preferred 42.3-day rotation period, although neither the periods nor their yearly aliases overlapped.  If the rotation period is instead closer to 20 days, it may be near that of the 22.8-day planet f discovered by V15, while the very long rotation period suggested by the M15 $v\sin i$ measurement is close to the 94.2 day period of V15's planet g.  Given the abundance and timespan of our \shk data, we sought to determine whether our observations could offer a more conclusive determination of the rotation period.

The presence of a large-amplitude activity cycle complicates the identification of a rotation period.  First, the rotation signal is superimposed over the larger magnetic cycle, thus requiring an adequate model of the cycle to reveal the residual rotation period.  Also, as observed for the Sun and other stars \citep{marchwinski15,diaz15}, the starspots/active regions which imprint the rotation period in the \shk/RV time series may appear and disappear during the maximum and minimum, respectively, of the activity cycle.  Finally, an insufficient observing cadence may introduce aliases that dwarf the true rotation period in frequency analysis tools such as periodograms \citep[e.g.][]{robertson15}.  We have therefore analyzed our \shk time series in a number of configurations in order to minimize the above complications and determine the most likely rotation period.

We started by examining only the 52 observations taken between September 2008 and January 2009, when the activity cycle passed through its maximum.  Restricting our analysis thusly serves two purposes: searching near the cycle maximum increases the likelihood that starspots or active regions will actually be present, and the essentially flat slope of the activity cycle in this region removes the need to model and remove the long-period signal, eliminating any systematics introduced in residual analysis.

\begin{figure}
\centering
\epsscale{1.275}
\plotone{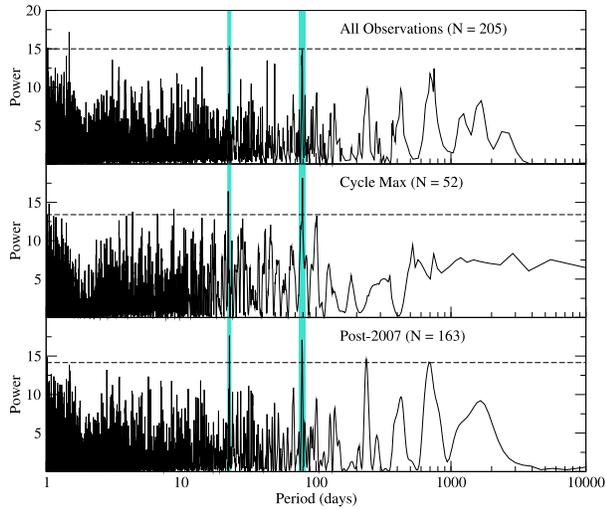}
\caption{Generalized Lomb-Scargle periodograms of \shk showing candidate rotation periods.  In addition to the full time series (top), we show subsets of the data from the peak of the activity cycle (middle), and the densely-sampled post-2007 observations (bottom). In all but the cycle max spectrum, we have removed the signal of the activity cycle.  Candidate periods at 22.8 and 79.4 days are highlighted.}
\label{fig:rot}
\end{figure}

In Figure \ref{fig:rot} (middle panel), we show the generalized Lomb-Scargle periodogram of the \shk observations from the cycle maximum.  There are two dominant peaks at periods of 22.4 and 79.4 days.  We note that the beat frequency between these periods is 32 days, leading us to suspect these peaks are monthly aliases of one another, and arise from the same physical origin. We verified this hypothesis by creating synthetic \shk datasets which sample the activity cycle and a superimposed sinusoid (plus white noise) at the epochs of our real observations.  In cases where the period of the superimposed sinusoid was either 22 or 79 days, peaks at both periods appeared in the periodograms.  This was especially evident when the simulated sinusoid was not constant in amplitude or phase, as we expect for a real signal due to stellar rotation.

The periodicities at 22 and 79 days remain prominent in our \shk observations regardless of how the data are subdivided or modeled. We note that for our further analysis we excluded the three spectra from 2009 Oct. 27 UT due to anomalously high $S_{HK}$ levels, possibly due to a flare event. In the top panel of Figure \ref{fig:rot}, we show the periodogram of our residual \shk measurements after modeling a sinusoid to the activity cycle.  We draw particular attention to the observations from 2007 to present (bottom panel of Figure \ref{fig:rot}), as our observing cadence for the star increased significantly, enabling greater sensitivity to the stellar rotation.  The pre-2007 data are also dominated by the activity cycle minimum, further reducing the likelihood of a robust detection of stellar rotation.  We consistently observed strong peaks at 22 and 79 days in all the periodograms, but particularly in the more densely sampled data.  The 22-day period tends to be stronger, especially after modeling and removing the signal of the activity cycle.

We conclude that the peak near 22 days is more likely to be the actual periodicity than the alias near 79 days.  
Furthermore, when modeling a sinusoid to the residual \shk values after removing the activity cycle, we found that models near 79 days left residual aliases near 31 and 235 days that were not present when using a 22.8-day model.  

To obtain a ``best-fit'' rotation model, we fit a sinusoid to the high-cadence post-2007 \shk data. 
This yielded a period of $22.83 \pm 0.03$ days with \shk amplitude $0.0077 \pm 0.0003$,
which matches the orbital period of HD 219134 f found by V15 to within one standard deviation.  This raises the possibility that the RV signal at 22.8 days is not caused by an exoplanet, but rather by stellar magnetic activity.  We do not explore further tests of this scenario here (e.g.~correlation with measured RVs), as a conclusive determination requires a comprehensive analysis of the combined HARPS-N/HIRES/APF/TS23 datasets, which is beyond the scope of this paper.  However, we strongly recommend such an analysis in the immediate future.

Is the 22.8 day period the stellar rotation period, however? It is inconsistent with the value of $v\sin i=0.4 \pm 0.5$ found by M15 from their $R=115,000$ HARPS-N spectra, and the 42.3 day periodicity that they found in their activity indicators. The 22.8 day periodicity is too long to be the first harmonic of the 42.3 day periodicity. If HD 219134 is differentially rotating, it is possible that the 42.3 day periodicity could be due to activity at fast-rotating low latitudes, and the 22.8 day periodicity could be the first harmonic due to activity at higher latitudes ($2\times22.8$ days $=45.6$ days). For a solar-like degree of differential rotation \citep[a differential rotation law $\omega=\omega_0-\omega_1\sin^2\phi$, with a differential rotation parameter  $\alpha=\omega_1/\omega_0$=0.20 from][where $\phi$ is the latitude]{ReinersSchmitt02}, if 42.3 days is the equatorial rotation period, then the rotation period would be 45.6 days at a latitude of $37^{\circ}$. This suggests that this scenario is plausible. Additional circumstantial evidence for this scenario is that a two-spot configuration, such as would generate a signal at the first harmonic, might be more likely near the activity cycle maximum, which dominates our rotation signal, and less likely near the cycle minimum, when the HARPS-N observations occurred. This would also help to explain why V15 detected the 22.8 day signal but M15 did not. Nontheless, the current data are insufficient to either confirm or reject this scenario. 

Further observations could help to test this scenario, and to conclusively determine whether the 22.8 day RV signal is indeed due to stellar activity rather than a planet. The activity level of HD 219134 is currently increasing, with the next maximum expected around late 2018. If the 22.8-day RV signal begins to appear in continuing HARPS-N observations as the activity level increases, this would be strong evidence for the stellar origin of this signal. In order to test for this possibility V15 split their dataset into three portions, and did recover the 22.8-day periodicity in all three subsets; however, they did not quote the significance level of the recovery or if the other parameters are consistent between the different subsets, preventing us from making a more detailed analysis of this issue. Additionally, high-cadence observations near the cycle maximum could potentially probe whether the 22.8 day periodicity is a harmonic of a longer rotation period.

If the 22.8 day RV signal is in fact caused by activity instead of a planet, it is very interesting in the context of the specific physical mechanism that creates the Doppler shifts.  Rather than examining periodicities and correlations in their residual \shk values, V15 sought to rule out a false-positive detection of HD 219134 f by obtaining photometry of the star.  HD 219134 is quiet photometrically, leading V15 to conclude that it must not exhibit large spots such as would create the RV signature at 22.8 days.  This test has proven hazardous in the case of M dwarfs, where activity signals have been observed to create RV signatures--sometimes mimicking exoplanets--with very little or no associated photometric variability \citep[e.g.][]{kurster03,robertson14}.  This phenomenon has not been observed for K dwarfs to date, but it would appear that HD 219134 is a candidate example.

M15 performed an analysis of their $\log R^\prime_{HK}$ data, and argued that their 42.3 day stellar rotation period was sufficiently well separated from their 46.78 day RV signal that this RV signal could not be a false positive due to stellar rotation. 
The 46.7 day period of HD 219134 d, however, is closer to twice the 22.8 day periodicity (45.6 days). 
Given our hypothesis on the origin of the 22.8 and 42.3 day activity signals, it is possible that activity at even higher latitudes could give rise to an RV signal at 46.7 days. No signal with this period is detected either in our activity data or that of M15, suggesting that 
 the 46.7 day RV signal is likely due to an actual planet, HD 219134 d.

\begin{figure}
\centering
\epsscale{1.25}
\plotone{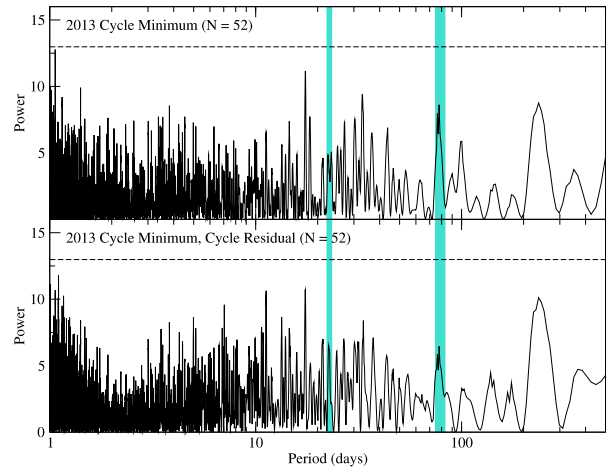}
\caption{Generalized Lomb-Scargle periodograms of \shk near the 2013 activity cycle minimum. For direct comparison to the activity cycle maximum (top middle panel of Fig.~\ref{fig:rot}) we use the same number of datapoints (52). In the top panel we show the periodogram of the \shk data, and in the bottom panel these same data after the best-fit activity cycle has been subtracted off. The vertical teal bars mark the candidate rotation periods of 22.8 and 79.4 days found earlier. No significant power is found at either periodicity in the minimum activity dataset.}
\label{fig:rot:min}
\end{figure}

The stellar rotation period also has consequences for the stellar age. \cite{Takeda07} found a stellar age of $12.5\pm0.5$ Gyr based upon a Bayesian isochrone analysis using stellar parameters derived from high signal-to-noise, high-resolution spectra. It is, however, very difficult to derive accurate isochrone ages for main sequence stars at this $T_{\mathrm{eff}}$, where isochrones for ages ranging from 1 to 12 Gyr differ by only $\sim0.1$ dex in $\log g$. 

Gyrochronology, on the other hand, gives a very different picture of the system age. Using the gyrochronological relation given by Eqn.~3 of \cite{Barnes07}, the rotation period of 42.3 days found by M15 implies an age of 4.1 Gyr. If 22.8 days were to be the rotation period, this would imply a gyrochronological age of 1.3 Gyr. Additionally, the activity-age relationship of \cite{MamajekHillenbrand08} predicts an age of 4.6 Gyr based upon our average activity level of $\log R'_{HK}=-4.89$. This relationship, however, is calibrated for F7-K2 dwarfs, and so is not strictly applicable to the K3V star HD 219134. Nonetheless, the stellar activity level is most consistent with a $\sim$40-day rotation period. 

Other aspects of the system could also shed light on its age. \cite{VolkGladman15} argued that systems of tightly-packed super-Earths are metastable over Gyr timescales, and eventually destabilize, resulting in the destruction of some of the planets. V15 did not perform a dynamical stability analysis, while M15 showed that their four-planet system is stable for at least $10^6$ orbits of HD 219134 e ($\sim5\times10^6$ years). Although it is beyond the scope of the present work, we suggest that a longer-term stability analysis of the HD 219134 system should be undertaken. If this shows that the system is only stable for a few Gyr, this would favor 22.8 days being the rotation period and a younger system age. 

\section{Radial Velocities}

\subsection{Radial Velocity Analysis}

\begin{figure*}
\centering
\plotone{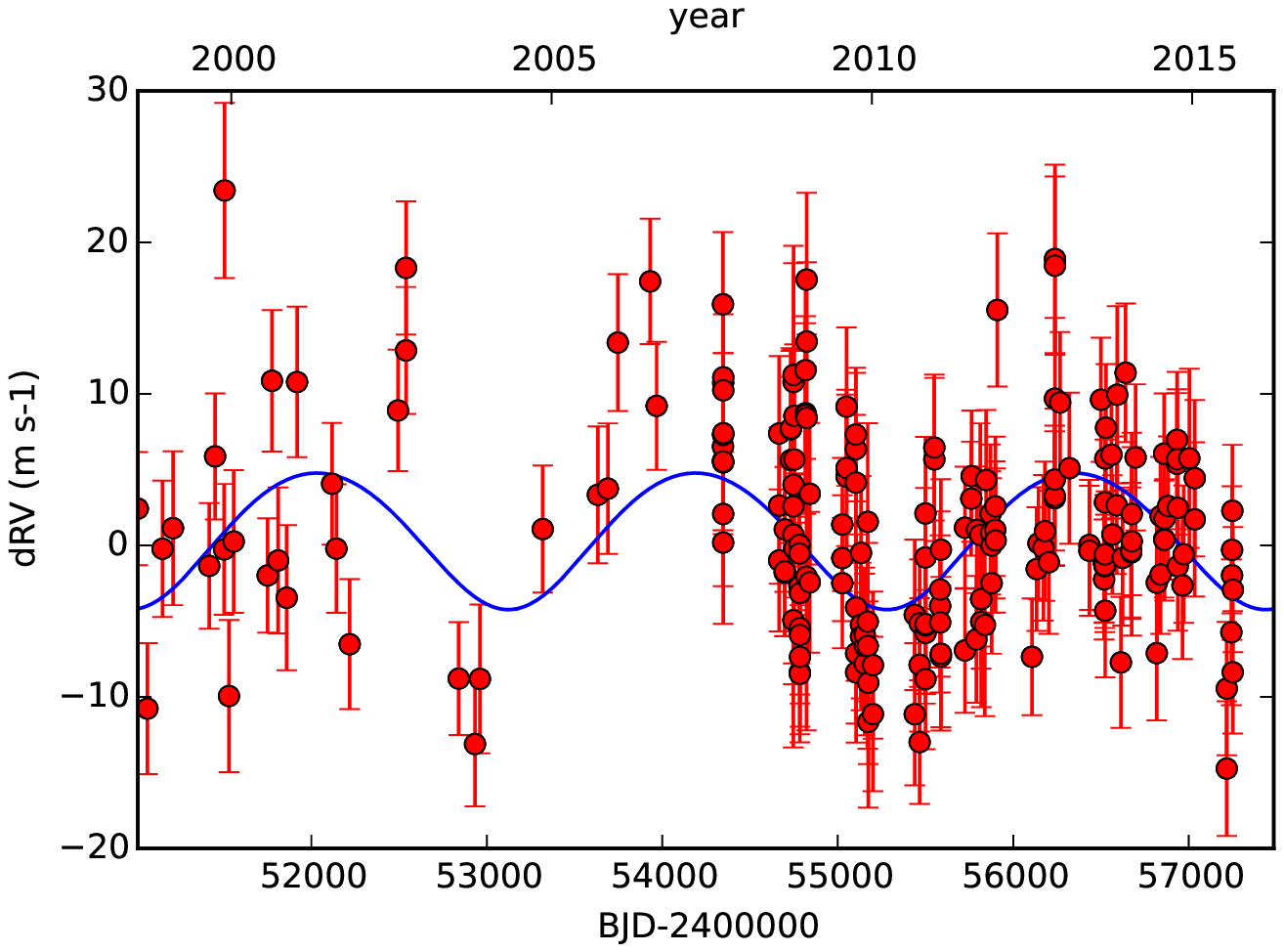}
\plotone{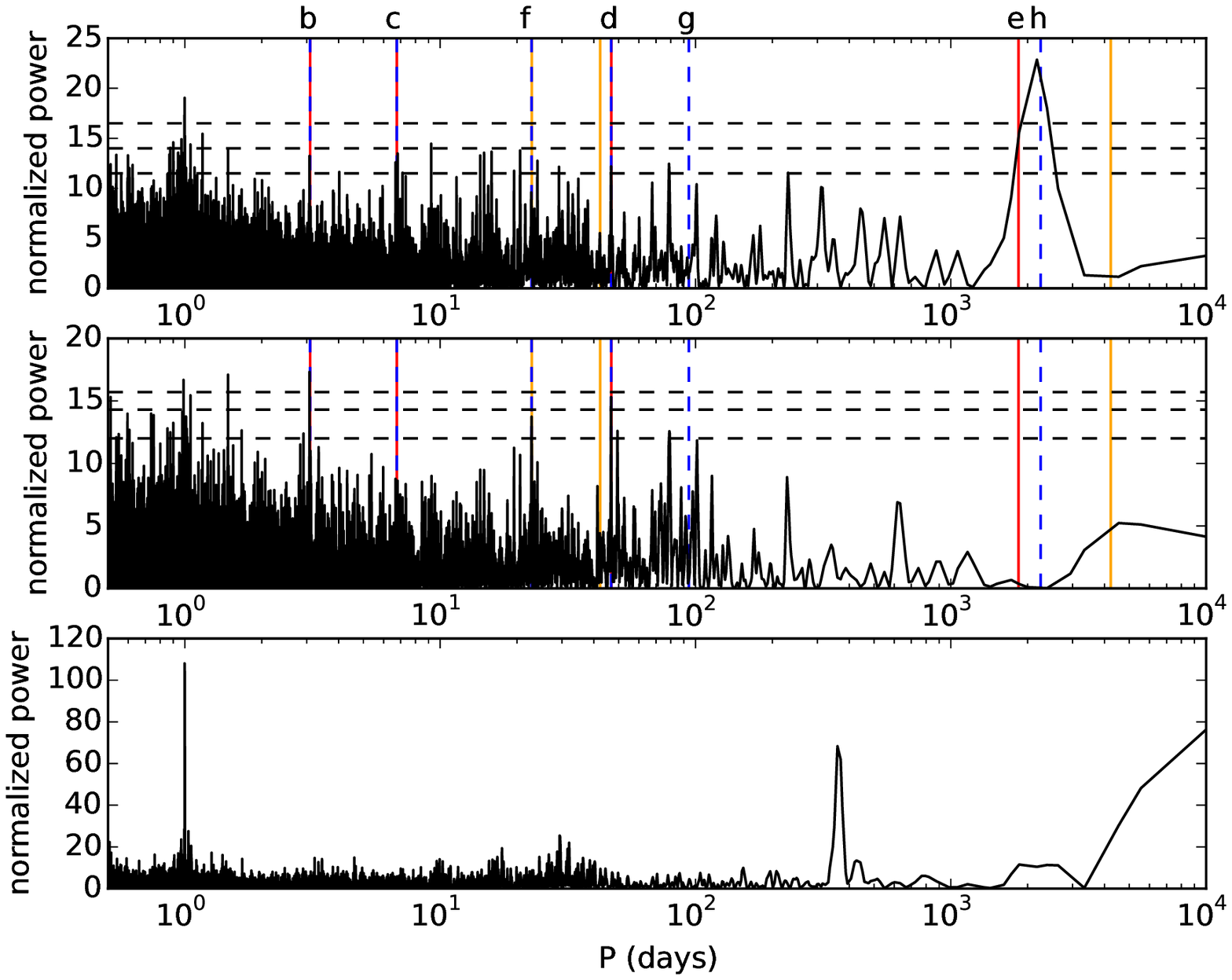}
\caption{Top panel: TS23 radial velocities of HD 219134, with the best-fit single planet model for HD 219134 h with eccentricity fixed to the value of $e=0.06$ from V15 overplotted in blue. Top middle: generalized Lomb-Scargle periodogram of the RVs. Vertical red lines show the periods of the planets found by M15, dashed blue lines show those of V15, and solid orange lines show, from left to right, our 22.8 day activity periodicity, the 42.3 day periodicity found by V15, and the activity cycle period that we found with our $S_{HK}$ measurements. Letters above the top axis show the names of the planets; note that M15 and V15 used different names for the long-period Saturn mass planet due to their disagreement about the orbital period, labeling it HD 219134 e and h, respectively. Horizontal dashed lines show bootstrapped FAP levels of, from top to bottom, 0.1\%, 1\%, and 10\%. Bottom middle: periodogram of the residuals after subtracting the best-fit RV signal of HD 219134 h (with $e=0.06$ fixed at the value of V15). Bottom: Window function for the TS23 RVs.}
\label{TS23RVs}
\end{figure*}

We used the Systemic Console 2 package\footnote{http://www.stefanom.org/systemic/} \citep{Meschiari09} to analyze our radial velocity data. 
We first analyzed only the TS23 Phase III data. We rejected all datapoints with internal uncertainties of $>8$ m s$^{-1}$ ($>3\sigma$ above the mean internal uncertainty of the dataset). These data are shown in the top panel of Fig.~\ref{TS23RVs}, and the corresponding Lomb-Scargle periodogram in the second panel of Fig.~\ref{TS23RVs}.

The periodogram shows a strong peak at $\sim2000$ days, with a bootstrapped false alarm probability (FAP) level of 0.02\%. This corresponds to the planet HD 219134 h found by V15. The 1842-day period of the outer planet HD 219134 e found by M15 lies away from this peak. We thus confirm the slightly longer period for this planet found by V15, although, again, the M15 value is very unconstrained.

When fitting to the TS23 data alone and ignoring the inner planets, we obtained a large eccentricity for HD 219134 h of $e=0.37 \pm 0.18$, in disagreement with the value of $e=0.06\pm0.04$ found by V15. This best-fit model, however, contained excursions to large negative velocities at epochs where we have no observations. We therefore fixed the eccentricity to the value found by V15. This resulted in only a slight increase in the reduced $\chi^2$, from $\chi^2_{\mathrm{red}}=1.63$ to $\chi^2_{\mathrm{red}}=1.69$. We also performed fits to the TS23 data alone, assuming the presence of three and five inner planets with parameters fixed at those found by M15 and V15. We again fixed the eccentricity of HD 219134 h to the value found by V15 and let its other parameters float. The results of this exercise are listed in Table~\ref{hpars}. Overall, the parameters that we found for HD 219134 h were broadly consistent, although they varied slightly depending upon the assumptions regarding the inner planets. We, however, obtained a slightly shorter period (2127-2198 days) and less massive (0.240-0.281 $M_J$) outer planet than V15 did; they found $P=2247 \pm 43$ days and $M\sin i=0.34 \pm 0.02 M_J$. Our results are formally also consistent with those of M15, but their observations covered approximately half of the orbit of HD 219134 e, and so they were unable to precisely measure its period (finding $P=1842_{-292}^{+4199}$ days). 

\begin{deluxetable*}{lccccc}
\tabletypesize{\scriptsize}
\tablecolumns{6}
\tablewidth{0pt}
\tablecaption{Parameters of the Outer Saturn-Mass Planet from McDonald Phase III Data \label{hpars}}
\tablehead{
\colhead{Parameter} & \colhead{1, $e$ free} & \colhead{1, $e$ fixed} & \colhead{1+5 V15} & \colhead{1+3 M15}
}

\startdata
$P$ (days) & $2198 \pm 51$ & $2146 \pm 64$ & $2127 \pm 63$ & $2121 \pm 61$ \\
$M\sin i$  $(M_J)$ & $0.281 \pm 0.056$ & $0.240 \pm 0.034$ & $0.256 \pm 0.029$ & $0.243 \pm 0.031$ \\
mean anomaly ($^{\circ}$) & $214 \pm 26$ & $351 \pm 66$ &  $260 \pm 75$ & $322 \pm 76$ \\
$e$ & $0.37 \pm 0.18$ & 0.06 (fixed) & 0.06 (fixed) & 0.06 (fixed) \\
$\omega$ ($^{\circ}$) & $180 \pm 19$ & $192 \pm 63$ & $106 \pm 73$ & $168 \pm 74$ \\
$a$ (AU) & $3.064 \pm 0.048$ & $3.015 \pm 0.060$ & $3.000 \pm 0.060$ & $2.975 \pm 0.057$ \\
$K$ (m s$^{-1}$) & $5.50 \pm 1.3$ & $4.42 \pm 0.62$ & $4.72 \pm 0.53$ & $4.54 \pm 0.58$ \\
$t_{\mathrm{peri}}$ (BJD) & 2448616.291 & 2448885.964 & 2448454.270 & 2455190.556\\
$\gamma$ (m s$^{-1}$) & $-1.04$ & $-0.27$ & $0.74$ & $-0.04$ \\
$\chi^2_{\mathrm{red}}$ & 1.63 & 1.69 & 1.59 & 1.70 
\enddata

\tablecomments{Best-fit parameters for the single-planet Systemic 2 fit to the TS23 data with various assumptions about the inner planets (see text for more details). Values quoted are median and mean absolute deviation values from a Systemic MCMC. We assumed a stellar mass of 0.794 $M_{\odot}$ from V15 for all fits except that with the M15 inner planet parameters, where we used the value of 0.78 $M_{\odot}$ from their work.}

\end{deluxetable*}

Detection of the super-Earths reported by M15 and V15 is very challenging, due to their small radial velocity semi-amplitudes (ranging from 1.1 to 4.4 m s$^{-1}$). Nonetheless, after subtracting the $\sim$2200-day signal from our McDonald Phase III data, we consistently recovered a 3.08 day signal as the strongest peak in the GLS periodogram of the residuals (third panel of Fig.~\ref{TS23RVs}), regardless of the details of the $\sim$2200-day signal. We fixed the eccentricity of HD 219134 h to the value of 0.06 found by V15, leaving the other parameters free, and subtracted the resulting best-fit signal from the TS23 data. The strongest signal in the periodogram of the residuals had a period of 3.08 days and a FAP of 0.02\%. There were also signals with periods of 46.7 days (6$^{\mathrm{th}}$ strongest signal, FAP=0.32\%), 22.8 days (10$^{\mathrm{th}}$ strongest signal, FAP=1.7\%), and 79.1 days (22$^{\mathrm{nd}}$ strongest signal, FAP=5.5\%). These signals correspond to HD 219134 d and f and the likely rotation alias seen in the $S_{HK}$ data (\S\ref{Paulswork}). While these were not the strongest signals in the periodogram of the residuals, they were three of the four strongest peaks with $P>2$ days (the fourth peak being the 21$^{\mathrm{st}}$ strongest peak, at 49.5 days and with a FAP of 5.5\%). Nonetheless, except for the 3.08-day signal we could not claim these as detections if we did not have prior knowledge of their existence. 
As the radial velocity precision of the TS23 data is significantly lower than those of the data published by M15 and V15, we did not pursue the analysis of the TS23 RV data alone any further. Nonetheless, this demonstrates that  despite the lower radial velocity precision of the Tull Spectrograph with respect to modern high-stability spectrographs like HARPS-N and APF, we are capable of detecting short-period super-Earths for bright stars with large amounts of data \citep[cf. 55 Cnc e:][]{McArthur04,Endl12}. 

\begin{figure}
\centering
\epsscale{1.25}
\plotone{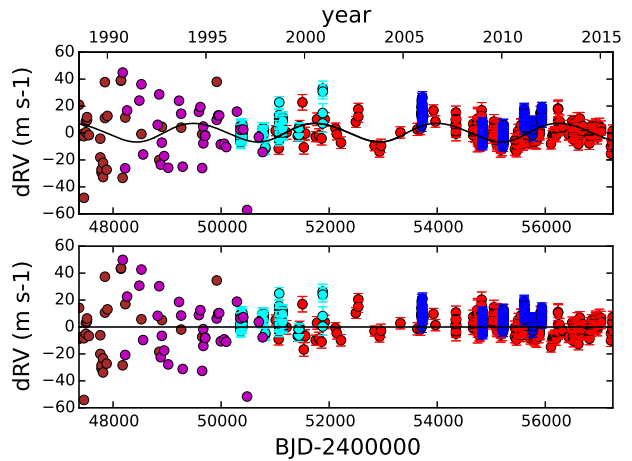}
\caption{Top panel: RVs from all five datasets: Phase I (brown), Phase II (purple), Phase III (red), Keck old CCD (light blue), and Keck new CCD (dark blue). The best-fit single-planet model to all five datasets, with fixed $e=0.06$ per V15, is overplotted. No error bars are shown for the Phase I or II data; see the note to Table~\ref{HJSTdata} for more information. Bottom panel: RV residuals after the subtraction of the best-fit model.}
\label{allrvs}
\end{figure}

We also attempted an analysis of our full five-part dataset (McDonald Phase I, II, and III, and Keck old and new CCDs; see Fig.~\ref{allrvs}). We recovered RV signals with periods of 2273 days (HD 219134 h) and 46.69 days (HD 219134 d), but failed to recover the other RV signals at a significant level; the very uneven time sampling of the Keck data is problematic for periodogram analysis. We therefore did not pursue the analysis of this combined dataset any further. 

\subsection{Limits on Long-Period Companions}

The overall architectures of planetary systems are of interest for constraining models of planet formation \citep[e.g.,][]{BatyginLaughlin15}. We thus used our long time baseline of RV observations to constrain the presence of additional planets in the system on wider orbits than HD 219134 h. In order to assess the detectability of such planets based on our Phase III dataset, we ran a large-scale simulation where we analyzed a batch of synthetic datasets generated by the planetary signals of two-planet systems. First, we fixed the parameters of one planet to the best-fit values of HD~219134~h, as derived by modeling the Keck, APF, and McDonald Phase III datasets, again using Systemic Console 2. Here we obtained $P=2247$ days, $M\sin i=0.32$, and $e=0.16$. The parameters of the hypothetical second planet were chosen on a uniform grid in $P_2$ (40 values), $K_2$ (40 values) and mean anomaly $\mathcal{M}_2$ (400 values), which generated a set of 640,000 planetary signals. $P_2$ spanned between 3370 days (1.5 $P_1$) and 6,273 days (the temporal span of the McDonald data), while $K_2$ spanned between 5 and 10~m~s$^{-1}$. For simplicity we assumed a circular orbit for the outer companion, and neglected the inner super-Earths found by M15 and V15; as these planets all have orbital periods much shorter than we were probing and small RV semi-amplitudes, they should not have a significant effect on the results. Each of the planetary signals was computed by sampling the RV response at the epochs of the Phase III dataset. Noise was subsequently added to each observation, based on a random scrambling of the residuals from the best-fit model.

\begin{figure}
\centering
\epsscale{1.25}
\hspace{-24pt}
\plotone{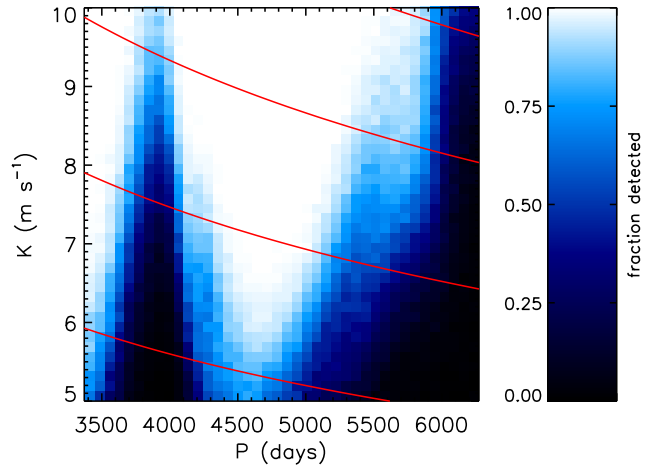}
\caption{Fraction of synthetic outer companions (beyond HD 219134 h) that were detected in our simulations; see the text for more details. The red contours show, from bottom to top, $M\sin i$ values of 0.375, 0.5, 0.625, and 0.75 $M_J$. We are less sensitive to hypothetical outer companions near the 2:1 and 3:1 resonances with HD 219134 h (at $P\sim4400$ and $6600$ days, respectively).}
\label{complimits}
\end{figure}

For each of the datasets, we ran a modeling procedure that fit any strong periodicities (FAP $<10^{-3}$) in the Lomb-Scargle periodogram of the data. We note that in most cases, only one of the planets was recovered. This is because the Phase III dataset is quite noisy compared to the semi-amplitude of HD~219134 h. The median formal uncertainty on the RV observations is $4.62$ m s$^{-1}$, and the RMS of the residuals for the best fit is $5.23$ m s$^{-1}$. Therefore, one or the other periodicity was often not evident in the residuals for 1-planet models. Figure \ref{complimits} shows the fraction of synthetic datasets for which the procedure either recovered both generated planets, or only the synthetic outer planet. For the latter point, we used the criterion that only the synthetic outer planet was detected if the recovered orbital period was closer to the input period of the hypothetical outer planet than to that of the inner planet HD 219134 h.

We computed companion mass limits at periods of 4340 days (the grid value most closely corresponding to the orbital period of Jupiter) and 5980 days (the longest period at which we can detect a significant number of synthetic companions with $K=5-10$ m s$^{-1}$) by finding the semi-amplitude below which, at that period, less than 68\% of synthetic signals were either detected as a second planet, or only the synthetic signal was detected. This excluded at $1\sigma$ companions with $M\sin i>0.36 M_J$ at 4340 days and $M\sin i>0.72 M_J$ at 5980 days. 
We thus demonstrate that, although HD 219134 possesses a Saturn-mass planet, there are likely no additional objects in the system with a mass approaching that of Jupiter and a period of less than 17 years. Such an object could lie on a near-face-on orbit, such that $\sin i$ is small, but as the innermost planet in the system transits, this would require a very large mutual inclination between such a hypothetical outer planet and at least one of the inner planets.

\subsection{Activity-Radial Velocity Correlation}
\label{arvcor}

\begin{figure}
\centering
\epsscale{1.25}
\plotone{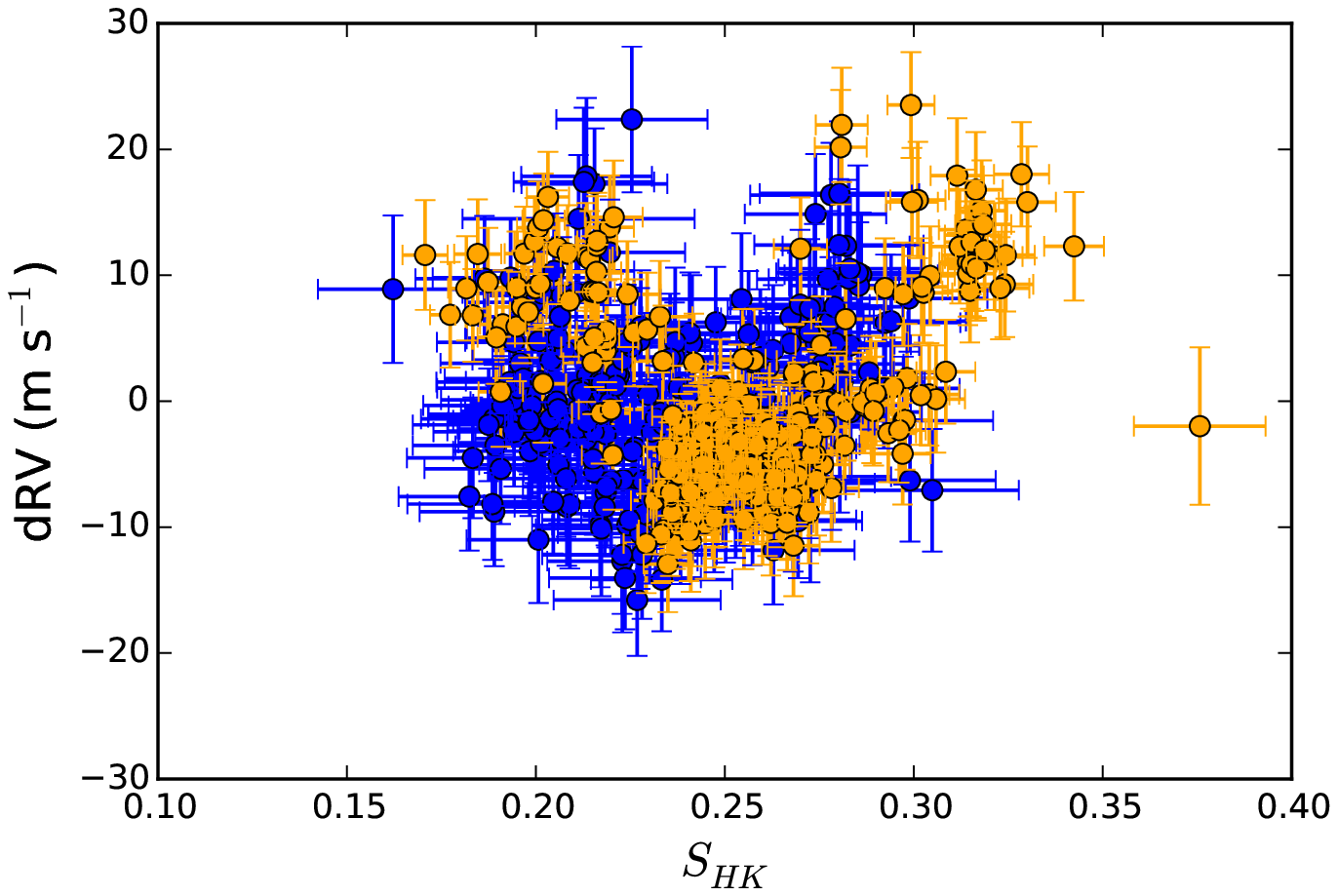}
\plotone{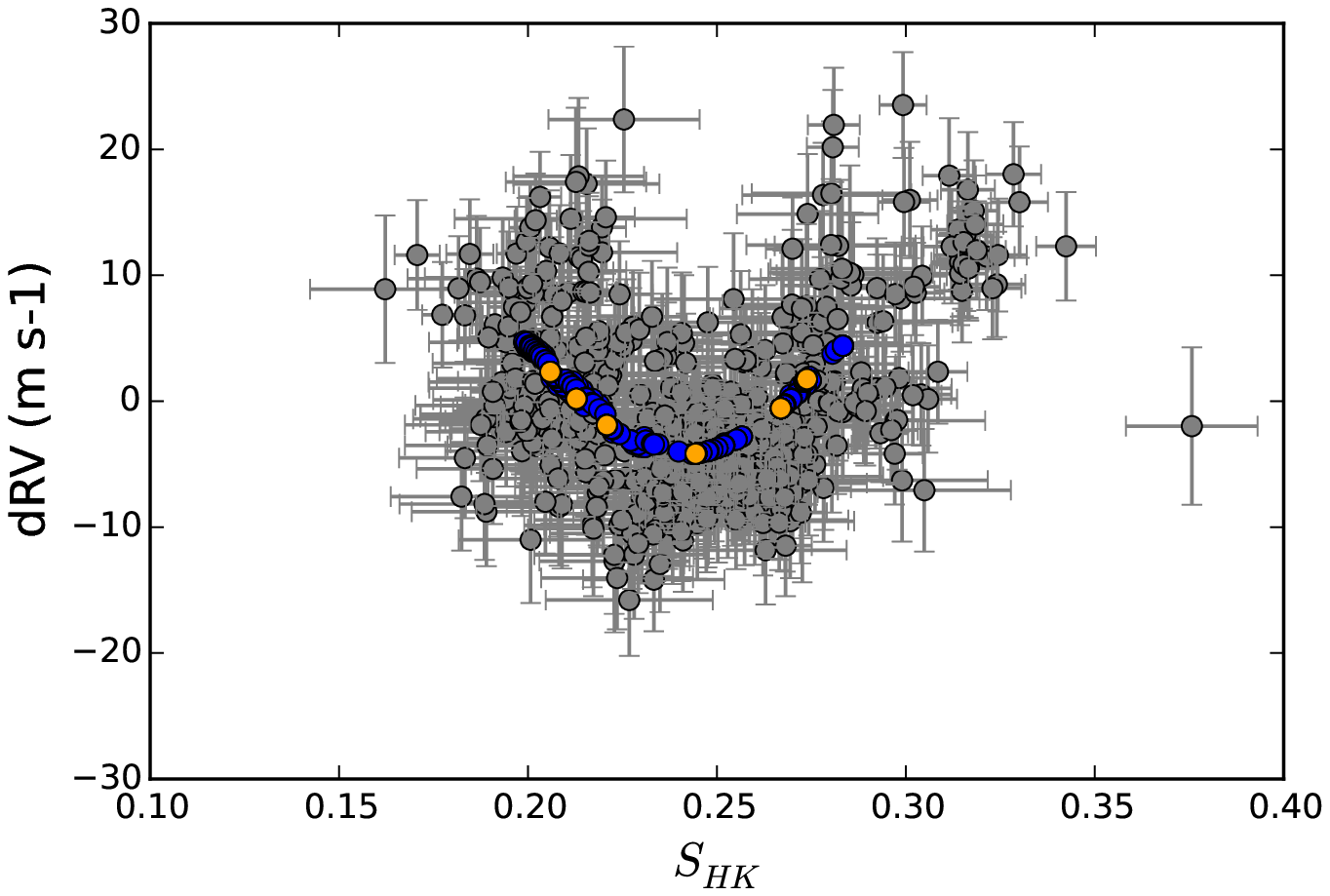}
\plotone{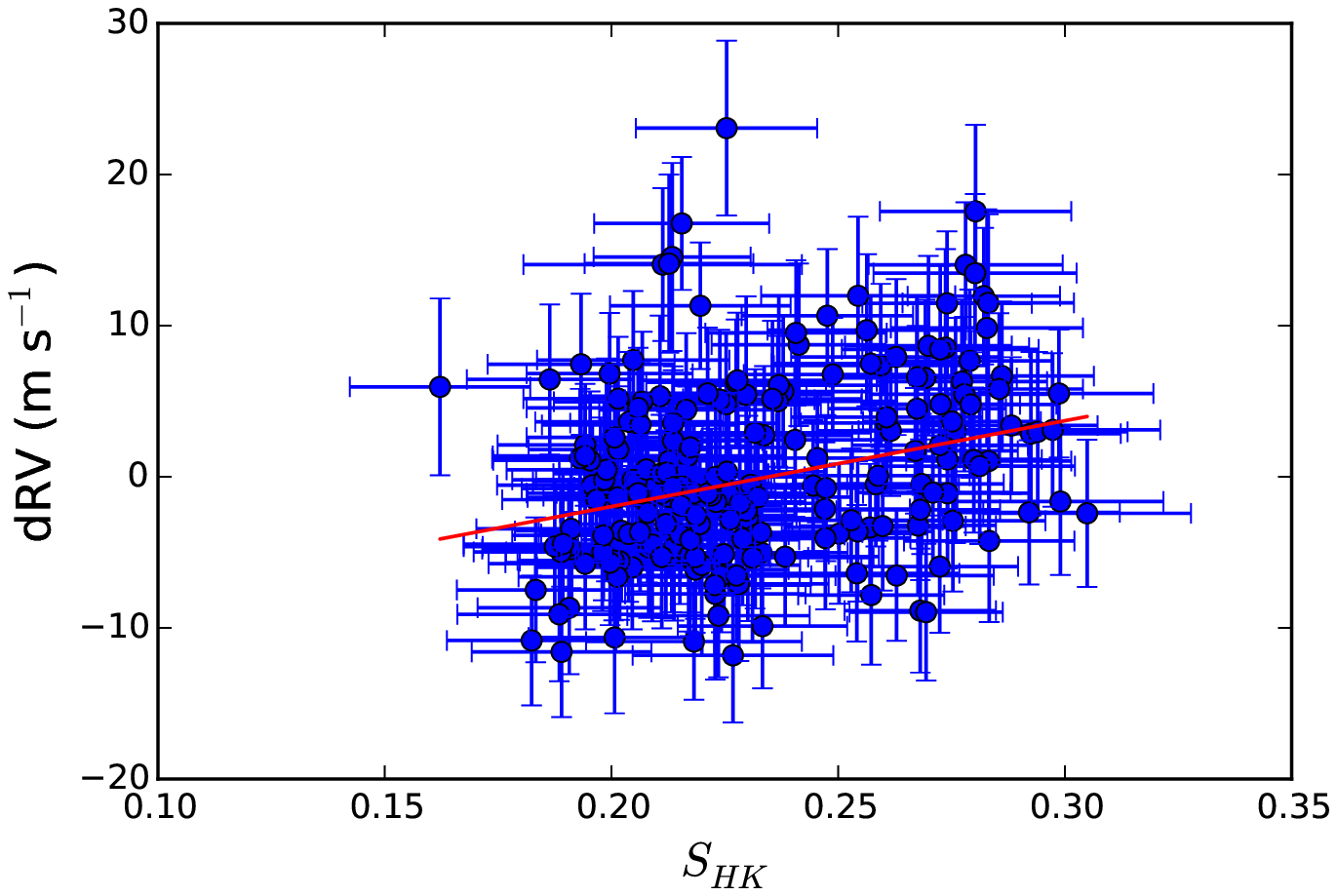}
\caption{Top panel: correlation between raw TS23 (blue) and new Keck CCD (orange) radial velocity and $S_{HK}$ measurements; an empirical multiplicative offset has been applied to the Keck \shk data to bring them into agreement with the sinusoidal trend seen in the TS23 data. Middle panel: correlation between the best-fit models of the activity cycle and the radial velocity of HD 219134 solely due to the outer planet HD 219134 h, for the epochs of the TS23 observations (blue) and new Keck CCD observations (orange), superimposed upon the data (gray). This model qualitatively reproduces the trend seen in the top panel. Bottom panel: correlation between the single-planet RV residuals and $S_{HK}$ measurements in the TS23 data. The Keck data are not shown because there is no trend visible. 
A red line shows the best linear fit to the data.}
\label{SMW_RV_COR}
\end{figure}

V15 found an unusual correlation between their $S_{HK}$ and RV measurements, with low RVs at moderate activity levels and increasing RV with both increasing {\it and} decreasing \shk (see Fig.~4 of their work). We show the correlation between our RVs and $S_{HK}$ measurements, before accounting for the presence of any planetary RV signals, in the top panel of Fig.~\ref{SMW_RV_COR}. We recover the same unusual RV-activity correlation found by V15.

A correlation of this form is puzzling. Previously observed RV-activity correlations have been linear \citep[e.g.,][]{Robertson13,Endl15}. This is expected theoretically because the radial velocity shifts are thought to be caused by magnetic fields associated with starspots suppressing surface convection locally, thus modifying the covering fraction of convective upwelling and downwelling on the stellar disk and shifting the observed radial velocity. 

We argue that the unusual form of the RV-activity correlation for HD 219134 is caused, in large part, simply by the near-commensurability of the orbital period of the Saturn-mass planet (2146 days, from our data) and the period of the activity cycle (4230 days). The activity period is close to twice that of the outer planet ($P_{SHK}/P_h=1.97 \pm 0.08$; or, instead using the period of planet h from V15, $P_{SHK}/P_h=1.88 \pm 0.07$). Over the past decade, these have conspired to align such that RV maxima due to HD 219134 h occur near the extrema of the activity cycle, while the RV minima occur at moderate activity levels. Such a pattern will naturally explain the correlation seen by V15 and in our own data, even in the absence of a causal relationship between the stellar activity and radial velocities. In the middle panel of Fig.~\ref{SMW_RV_COR} we show the correlation between the best-fit sinusoidal model for the TS23 $S_{HK}$ measurements and the best-fit single-planet model for HD 219134 h at the TS23 and Keck (new CCD only) measurement epochs. The beating of the two frequencies against each other produces a Lissajous figure; these models do qualitatively reproduce the observed trend of low RVs at moderate $S_{HK}$ values and high RVs at both low and high $S_{HK}$ values.

In order to test whether there could still be an actual correlation between the RVs and the activity cycle for HD 219134, we subtracted the single-planet best-fit model for HD 219134 h off of the RVs and searched for a correlation of the RV residuals with the corresponding $S_{HK}$ measurements. This is shown in the bottom panel of Fig.~\ref{SMW_RV_COR}. While there does not appear to be a trend in the new Keck data, a possible trend is evident in the TS23 Phase III data. 

We explored this trend by using the Pearson correlation test on the distribution of TS23 $S_{HK}$ measurements and RV residuals. We found a Pearson correlation coefficient of 0.30, which, for a sample size of 202 datapoints, corresponds to a p-value of $1\times10^{-5}$. This suggests that we are indeed seeing a genuine correlation, although this is not as statistically significant as those found by \cite{Endl15} for $\beta$ Vir and HD 10086.

\section{Conclusions}

HD 219134 is of substantial scientific interest as one of the nearest planet-host stars, and the nearest and brightest star known to host a transiting planet (M15, V15). We have presented 27 years' worth of McDonald Observatory radial velocity observations of this star, including 17 years of high-quality data utilizing the full iodine cell spectral bandpass. Using these data, we have detected the 6-year Saturn-mass planet HD 219134 h, and measured planetary parameters in broad agreement with those presented by V15, as well as the less precise values found by M15. 
Additionally, we have detected the low-amplitude RV signal due to the 3.08-day transiting super-Earth HD 219134 b, and have tentative (i.e., low-significance) detections of the RV signals at 22.8 and 46.7 days, corresponding to planets HD 219134 f and d.

We have also investigated the stellar activity of HD 219134, as measured from the Ca~\textsc{ii} H and K lines and quantified as \shk. We have detected a long-period activity cycle for the star, with a period of $4230 \pm 100$ days, very similar in length to our own Sun's activity cycle. Furthermore, by analyzing the \shk residuals after the subtraction of the activity cycle signal, we have found a significant periodicity at $22.83 \pm 0.03$ days, which we suggest may be the first harmonic due to activity at moderate latitude on a differentially rotating stellar surface.  This period is identical within the errors to the orbital period of the planet HD 219134 f found by V15. This suggests that the RV signal attributed to planet f may be a false positive due to the stellar rotation. Conversely, however, our work provides evidence that the 46.7 day HD 219134~d and the 2200 day HD 219134 h are likely to be actual planets; M15 and V15, respectively, had expressed some concerns about whether these RV signals could be related to stellar rotation or activity. This highlights the importance of accounting for activity variations due to both stellar rotation and long-term activity cycles for high-precision RV work, especially as the exoplanet community pushes to detect ever smaller RV variations.

HD 219134 is additionally important as the nearest and brightest star known to host an analog to the systems of closely-packed, short-period super-Earths found in great abundance by {\it Kepler}. The overall orbital architectures of such systems are of great interest to constrain how these systems formed \citep[e.g.,][]{BatyginLaughlin15}. These architectures can be probed through the combination of high-precision RVs and photometry to find small short-period planets, and long-term radial velocity observations to detect Jupiter and Saturn analogs. The McDonald Observatory Planet Search is in a strong position to provide this latter dataset, with its sample of long-term RV observations of more than 200 bright FGK stars approaching an observing baseline of 15 years. 

\vspace{12pt}

We thank the other observers who gathered some of our observations, including Diane Paulson, Kevin Gullikson, Stuart Barnes, Candace Gray, Anita Cochran, Suzanne Hawley, Ed Barker, and Elizabeth Ambrose. We thank Rachel Akeson and the NASA Exoplanet Archive team for providing recommendations on resolving the HD 219134 e/h naming ambiguity, and Steven Vogt for clarifying the reasoning behind the naming scheme used in V15.

This work has been made possible through the National Science Foundation (Astrophysics grant AST-1313075) and various NASA grants over the years. We
are grateful for their generous support. We also thank
the McDonald Observatory Time Allocation committee
for its continuing support of this program. S.M. acknowledges support from the W.~J. McDonald Postdoctoral Fellowship.

This paper includes data taken at The McDonald Observatory of The University of Texas at Austin. The authors acknowledge the Texas Advanced Computing Center (TACC, http://www.tacc.utexas.edu) at The University of Texas at Austin for providing HPC resources that have contributed to the research results reported within this paper. Some of the data presented herein were obtained at the W.M. Keck Observatory, which is operated as a scientific partnership among the California Institute of Technology, the University of California and the National Aeronautics and Space Administration. The Observatory was made possible by the generous financial support of the W.M. Keck Foundation. The Keck observations were partially obtained through the CoRoT NASA Key Science Project allocation. The authors wish to recognize and acknowledge the very significant cultural role and reverence that the summit of Mauna Kea has always had within the indigenous Hawaiian community.  We are most fortunate to have the opportunity to conduct observations from this mountain.

{\it Facilities:} McDonald Observatory (Tull coud\'e spectrograph), Keck-I Observatory (HIRES)


\begin{thebibliography}{}

\bibitem[Artigau et al.(2014)]{Artigau14} Artigau, {\'E}., 
Kouach, D., Donati, J.-F., et al.\ 2014, \procspie, 9147, 914715 

\bibitem[Baliunas et al.(1995)]{Baliunas95} Baliunas, S.~L., 
Donahue, R.~A., Soon, W.~H., et al.\ 1995, \apj, 438, 269 

\bibitem[Barnes(2007)]{Barnes07} Barnes, S.~A.\ 2007, \apj, 669, 1167 

\bibitem[Batygin \& Laughlin(2015)]{BatyginLaughlin15} Batygin, K., \& Laughlin, G.\ 2015, PNAS, 112, 4214 

\bibitem[Cochran \& Hatzes(1993)]{CochranHatzes93} Cochran, W.~D., \& Hatzes, A.~P.\ 1993, in Planets Around Pulsars, ed. J.~A. Phillips, J.~E. Thorsett, \& S.~R. Kulkarni, 36, 267 

\bibitem[Cochran et al.(2002)]{Cochran02} Cochran, W.~D., Hatzes, 
A.~P., \& Paulson, D.~B.\ 2002, \aj, 124, 565 

\bibitem[Coughlin et al.(2015)]{Coughlin15} Coughlin, J.~L., 
Mullally, F., Thompson, S.~E., et al.\ 2015, arXiv:1512.06149 

\bibitem[D{\'{\i}}az et al.(2016)]{diaz15} D{\'{\i}}az, R.~F., S{\'e}gransan, D., Udry, S., et al.\ 2016, \aap, 585, A134 

\bibitem[Endl et al.(2016)]{Endl15} Endl, M., Brugamyer, 
E.~J., Cochran, W.~D., et al.\ 2016, \apj, 818, 34 

\bibitem[Endl et al.(2000)]{Endl00} Endl, M., K{\"u}rster, M., \& Els, S.\ 2000, \aap, 362, 585 

\bibitem[Endl et al.(2012)]{Endl12} Endl, M., Robertson, P., 
Cochran, W.~D., et al.\ 2012, \apj, 759, 19 

\bibitem[Foreman-Mackey et al.(2013)]{emcee} Foreman-Mackey, D., Hogg, D.~W., Lang, D., \& Goodman, J.\ 2013, \pasp, 125, 306 

\bibitem[Hatzes \& Cochran(1993)]{HatzesCochran93} Hatzes, A.~P., \& Cochran, W.~D.\ 1993, \apj, 413, 339 

\bibitem[Hatzes et al.(2003)]{Hatzes03} Hatzes, A.~P., Cochran, 
W.~D., Endl, M., et al.\ 2003, \apj, 599, 1383 

\bibitem[Jones et al.(2010)]{Jones10} Jones, H.~R.~A., Butler, 
R.~P., Tinney, C.~G., et al.\ 2010, \mnras, 403, 1703 

\bibitem[K\"urster et al.(2003)]{kurster03} K\"urster, M., Endl, M., Rouesnel, F., et al.\ 2003, \aap, 403, 1077

\bibitem[Lovis et al.(2011)]{Lovis11} Lovis, C., Dumusque, X., 
Santos, N.~C., et al.\ 2011, arXiv:1107.5325 

\bibitem[Mamajek \& Hillenbrand(2008)]{MamajekHillenbrand08} Mamajek, E.~E., \& Hillenbrand, L.~A.\ 2008, \apj, 687, 1264 

\bibitem[Marchwinski et al.(2015)]{marchwinski15} Marchwinski, R.~C., Mahadevan, S., Robertson, P., Ramsey, L., \& Harder, J.\ 2015, \apj, 798, 63

\bibitem[Marmier et al.(2013)]{Marmier13} Marmier, M., S{\'e}gransan, D., Udry, S., et al.\ 2013, \aap, 551, A90 


\bibitem[McArthur et al.(2004)]{McArthur04} McArthur, B.~E., Endl, 
M., Cochran, W.~D., et al.\ 2004, \apjl, 614, L81 

\bibitem[M{\'e}gevand et al.(2014)]{Megevand14} M{\'e}gevand, D., 
Zerbi, F.~M., Di Marcantonio, P., et al.\ 2014, \procspie, 9147, 91471H 


\bibitem[Meschiari et al.(2009)]{Meschiari09} Meschiari, S., Wolf, 
A.~S., Rivera, E., et al.\ 2009, \pasp, 121, 1016 

\bibitem[Motalebi et al.(2015)]{Motalebi15} Motalebi, F., Udry, S., Gillon, M., et al.\ 2015, \aap, 584, A72 


\bibitem[Paulson et al.(2002)]{Paulson02} Paulson, D.~B., Saar, 
S.~H., Cochran, W.~D., \& Hatzes, A.~P.\ 2002, \aj, 124, 572 

\bibitem[Perryman et al.(2014)]{Perryman14} Perryman, M., Hartman, 
J., Bakos, G.~{\'A}., \& Lindegren, L.\ 2014, \apj, 797, 14 

\bibitem[Quirrenbach et al.(2014)]{Quirrenbach14} Quirrenbach, A., 
Amado, P.~J., Caballero, J.~A., et al.\ 2014, \procspie, 9147, 91471F 

\bibitem[Reiners \& Schmitt(2002)]{ReinersSchmitt02} Reiners, A., \& Schmitt, J.~H.~M.~M.\ 2002, \aap, 384, 155 

\bibitem[Ricker et al.(2015)]{Ricker15} Ricker, G.~R., Winn, 
J.~N., Vanderspek, R., et al.\ 2015, JATIS, 1, 014003 

\bibitem[Robertson et al.(2013)]{Robertson13Mact} Robertson, P., Endl, 
M., Cochran, W.~D., \& Dodson-Robinson, S.~E.\ 2013, \apj, 764, 3 

\bibitem[Robertson et al.(2013)]{Robertson13} Robertson, P., Endl, 
M., Cochran, W.~D., MacQueen, P.~J., \& Boss, A.~P.\ 2013, \apj, 774, 147 

\bibitem[Robertson et al.(2014)]{robertson14} Robertson, P., Mahadevan, S., Endl, M., \& Roy, A.\ 2014, Science, 345, 440

\bibitem[Robertson et al.(2015)]{robertson15} Robertson, P., Roy, A., \& Mahadevan, S.\ 2015, \apjl, 805, L22

\bibitem[Rowan et al.(2016)]{Rowan15} Rowan, D., Meschiari, S., 
Laughlin, G., et al.\ 2016, \apj, 817, 104 

\bibitem[Rowe et al.(2014)]{Rowe14} Rowe, J.~F., Bryson, 
S.~T., Marcy, G.~W., et al.\ 2014, \apj, 784, 45 

\bibitem[Soderblom et al.(1991)]{Soderblom91} Soderblom, D.~R., 
Duncan, D.~K., \& Johnson, D.~R.~H.\ 1991, \apj, 375, 722 

\bibitem[Swift et al.(2015)]{Swift15} Swift, J.~J., Bottom, M., 
Johnson, J.~A., et al.\ 2015, JATIS, 1, 027002 

\bibitem[Takeda et al.(2007)]{Takeda07} Takeda, G., Ford, E.~B., 
Sills, A., et al.\ 2007, \apjs, 168, 297 

\bibitem[Tokovinin et al.(2013)]{Tokovinin13} Tokovinin, A., 
Fischer, D.~A., Bonati, M., et al.\ 2013, \pasp, 125, 1336 

\bibitem[Tull et al.(1995)]{Tull95} Tull, R.~G., MacQueen, 
P.~J., Sneden, C., \& Lambert, D.~L.\ 1995, \pasp, 107, 251 

\bibitem[Valenti \& Fischer(2005)]{ValentiFischer05} Valenti, J.~A., \& Fischer, D.~A.\ 2005, \apjs, 159, 141 

\bibitem[Vogt et al.(1994)]{Vogt94} Vogt, S.~S., Allen, S.~L., 
Bigelow, B.~C., et al.\ 1994, \procspie, 2198, 362 

\bibitem[Vogt et al.(2015)]{Vogt15} Vogt, S.~S., Burt, J., 
Meschiari, S., et al.\ 2015, \apj, 814, 12 

\bibitem[Vogt et al.(2014)]{Vogt14} Vogt, S.~S., Radovan, M., 
Kibrick, R., et al.\ 2014, \pasp, 126, 359 

\bibitem[Volk \& Gladman(2015)]{VolkGladman15} Volk, K., \& Gladman, B.\ 2015, \apjl, 806, L26 

\bibitem[Walker et al.(1995)]{Walker95} Walker, G.~A.~H., 
Walker, A.~R., Irwin, A.~W., et al.\ 1995, \icarus, 116, 359 

\bibitem[Zechmeister et al.(2013)]{Zechmeister13} Zechmeister, M., K{\"u}rster, M., Endl, M., et al.\ 2013, \aap, 552, A78 


\end{thebibliography}
\end{document}